\documentclass[twocolumn,showpacs,preprintnumbers,amsmath,amssymb]{revtex4}
\usepackage{graphicx}
\usepackage{dcolumn}
\usepackage{bm}
\begin {document} 
\title{Coexistence of open and closed gallery spaces in intercalation compounds}
\author{Ralf Br\"uning}
\email{rbruening@mta.ca}
\homepage{http://www.mta.ca/
\author{Steve Emeneau}
\author{Kristopher Bulmer}
\affiliation{Physics Department, Mount Allison University, Sackville, New Brunswick,
Canada E4L 1E6}
\author{Rabin Bissessur}
\author{Robert Haines}
\affiliation{Department of Chemistry, University of Prince Edward Island, Charlottetown, Prince Edward Island, Canada C1A 4P3}
\date{\today}

\begin{abstract}
A series of molybdenum disulfide intercalation compounds was prepared to study the effect
of varying loadings of the samples with intercalated molecules.  The molecules,
(5,7,7,12,14,14-hexamethyl-1,4,8,11-tetraazacyclotetradeca-4,11-diene) macrocycles,
occupy the gallery spaces between the MoS$_2$ layers in a single layer.
The samples were characterized by thermogravimetry, wide angle X-ray scattering,
scanning electron microscopy and transmission electron microscopy (TEM).
TEM images reveal the layer stacking as the filling of the gallery spaces with
macrocycles increases gradually up to one monolayer per gallery space.
Fourier transforms of these images are in
excellent agreement with the X-ray scattering of the bulk compound.
X-ray scattering by these compounds was simulated
by evaluating the Debye sum for weighted averages of model particles that have sequences
of open and closed gallery spaces.
Open and closed gallery spaces coexist in
the samples with less than one monolayer of intercalated molecules
per gallery space.  The fraction of open gallery spaces increases with macrocycle loading,
from zero for restacked MoS$_2$ without macrocycles to one for a fully intercalated compound.
The sequence of open and closed spaces is approximately random, and
the MoS$_2$ layers stack with the same orientation but with random lateral shifts.  The binding
enthalpy of the macrocycles in the gallery spaces is about 14 meV.

\end{abstract}

\maketitle

\section{Introduction}
\label{intro}
Intercalation compounds  are being developed for a wide range of purposes.
Molybdenum disulfide (MoS$_2$) intercalation compounds in particular,
with either organic \cite{div,yan1,bis2,bis3} or inorganic \cite{sch,pae} guest molecules inserted
between gallery spaces formed by the MoS$_2$ sheets,
have been researched extensively in recent years \cite{ben}.
This is partially driven by the use of MoS$_2$ as  a sulfur removal catalyst \cite{alo},
and its promise as
cathode materials in rechargeable batteries \cite{gon},  as a photocatalyst
and as a solar cell material \cite{pae}. 

In this paper we describe the properties of a series of intercalation compounds.  The
amount of intercalated organic macrocyclic molecules between MoS$_2$ sheets is varied
from zero to approximately maximal loading.  The macrocycles intercalate in a
single layer with the plane of the ring molecule parallel to the MoS$_2$ sheets, and
compounds with on average
less than a full intercalated monolayer per sheet are of particular interest.
The range of compounds is analyzed with
a combination of TEM imaging and the analysis of the X-ray scattering.
The X-ray scattering intensities are simulated with a method had been used
by Yang et al.\ to find the structure of
exfoliated MoS$_2$ sheets suspended in aqueous solution \cite{yan,gor}, and
this technique has recently been extended to find the local structure of a fully intercalated
macrocycle intercalation compound \cite{bru1}.
This analysis of the X-ray scattering of the macrocycle intercalation compounds
has shown that the MoS$_2$ layers stack with the same orientation, but with random
lateral translations between the layers.
Further, the positions of the atoms within
the MoS$_2$ layers have a large degree of static disorder.  This static disorder may be attributed to the random field originating from the guest molecules, whose effect is enhanced
as there are several competing intra-layer MoS$_2$ structures \cite{yan,hei,ben}.
High resolution transmission electron microscopy imaging of TiO$_2$--pillared MoS$_2$
shows that the layered structure has a wavy pattern (mosaic) with correlation lengths of about 10 nm
parallel and perpendicular to the stacking direction \cite{pae}.  The observed diffraction
peak widths would correspond to particle dimensions of about 10 nm.  Therefore the mosaic correlation
length, rather than the true particles size, governs the peak widths of the X-ray
diffraction pattern.
The MoS$_2$ layers are seen by scanning electron microscopy to be several
micrometers wide \cite{pae}.

\section{Experimental Methods}
Deuterated macrocycles, shown in figure \ref{macrocycle},
were prepared according to the literature \cite{dou}.
In order to carry out neutron scattering measurements on the intercalation compounds
that highlight the macrocycles relative to the MoS$_2$ sheets, the molecules were synthesized
with deuterium atoms in all but two of the hydrogen positions.  For the same reason the
intercalation was carried
out in deuterated water. The analysis of these neutron scattering measurements
will be subject of a forthcoming paper.
Li$_x$MoS$_2$ was prepared  as described in ref.\ \cite{bis3}.
Reaction of Li$_x$MoS$_2$ with D$_2$O results in the formation of completely
exfoliated single layers of MoS$_2$.
The intercalation compounds were synthesized by adding a solution of the
macrocycles in D$_2$O to the exfoliated MoS$_2$ solution.
The reaction mixture was allowed to stir at room temperature for two days.  The
mixture was then filtered and the insoluble product washed throroughly with D$_2$O to
remove LiOH and excess macrocycles.
\begin{figure}[b]
\begin{center}
\includegraphics{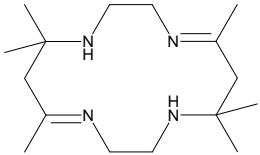}
\caption{\label{macrocycle}
Structure of the macrocycle molecule,
5,7,7,12,14,14-hexamethyl-1,4,8,11-tetraazacyclotetradeca-4,11-diene, deuterated at all C atoms.
}
\end{center}
\end{figure}

The macrocycle and D$_2$O  content of the intercalation compounds
was determined by thermogravimetric analysis (TGA) using a TA 500 instrument.
This method has been previously shown to give
results for the stoichiometry of the samples that agree closely with elemental analysis \cite{bis4}.
The sample mass used was typically between 5 and 10 mg, and samples
mass change was recorded upon heating the samples at 10 K/min in air.
The morphology of the compounds was determined with scanning electron microscope (SEM)
using a
JEOL JSM-5600 operated at 10 kV.  Prior to imaging, the particles were spread on glass
substrates and a 15 nm average thickness gold coating was applied to the
surface by plasma sputtering.
Transmission electron microscopy images were obtained with a JEOL 2011 at an
accelerating voltage of 200 kV.  TEM samples were prepared by embedding the
sample powders in LR White Resin.  Sections with a thickness of approximately
60 nm were prepared with a Leica Ultramicrotome and placed on a 200 mesh
copper grid.
The X-ray scattering of the samples was measured with a custom built
 $\theta$-$\theta$
 diffractometer. This instrument is equipped with graphite monochromator and analyzer
 crystals, and Cu-K$_{\alpha}$ radiation was used.  The sample powders were measured
 at room temperature on a (510)
 single crystal silicon substrate with negligible scattering in the direction
 perpendicular to the surface, and the
 sample chamber was evacuated to avoid air scattering.  The data are displayed as a function of the
 length of the scattering vector, $q = (4 \pi / \lambda) \sin \theta$.

\section{Results}

\label{results}
The mass of the intercalation compounds upon heating in air is plotted in figure
\ref{tga}.  The loading of the macrocycles, $r$, and the water content, $z$
of the compounds (D$_2$O)$_z$(macrocycle)$_r$MoS$_2$ is obtained
from these thermogravimetric traces.  Below 100$^{\circ}$C the weight change is dominated
by the evaporation of water.  In the range between 100$^{\circ}$C and 350$^{\circ}$C
the macrocycles evaporate, and above 350$^{\circ}$ MoS$_2$ decomposes and
oxidizes to
form MoO$_3$ at around 525$^{\circ}$C.  The initial mass of 
MoS$_2$, calculated based on the MoO$_3$ mass, is indicated
by the horizontal line in figure \ref{tga}.
The remainder of the initial mass at room temperature is due to residual D$_2$O and the
macrocycles.   The amount of macrocycles is determined by extrapolating the curved lines,
observed between about 100 and 400$^{\circ}$C, back to room temperature.  The remaining
initial mass is attributed to the residual water.  The macrocycle-free sample (lowest curve)
shows an additional mass loss process already below 100$^{\circ}$C that may be related to a
MoS$_2$ decomposition reaction.
The sample compositions found by this method are given in figure \ref{x-ray}.

\begin{figure}[t]
\begin{center}
\includegraphics{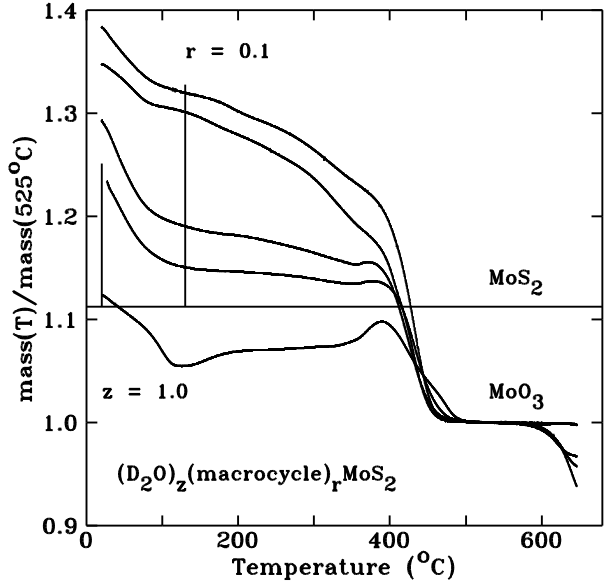}
\caption{\label{tga} Mass change of samples in the TGA, normalized to
the residual mass at 525C.
The lowest curve was obtained for a sample without macrcocycles.
Vertical bars show the expected weight changes for $r=0.1$ and $z=1$.}
\end{center}
\end{figure}

\begin{figure}[t]
\begin{center}
\includegraphics{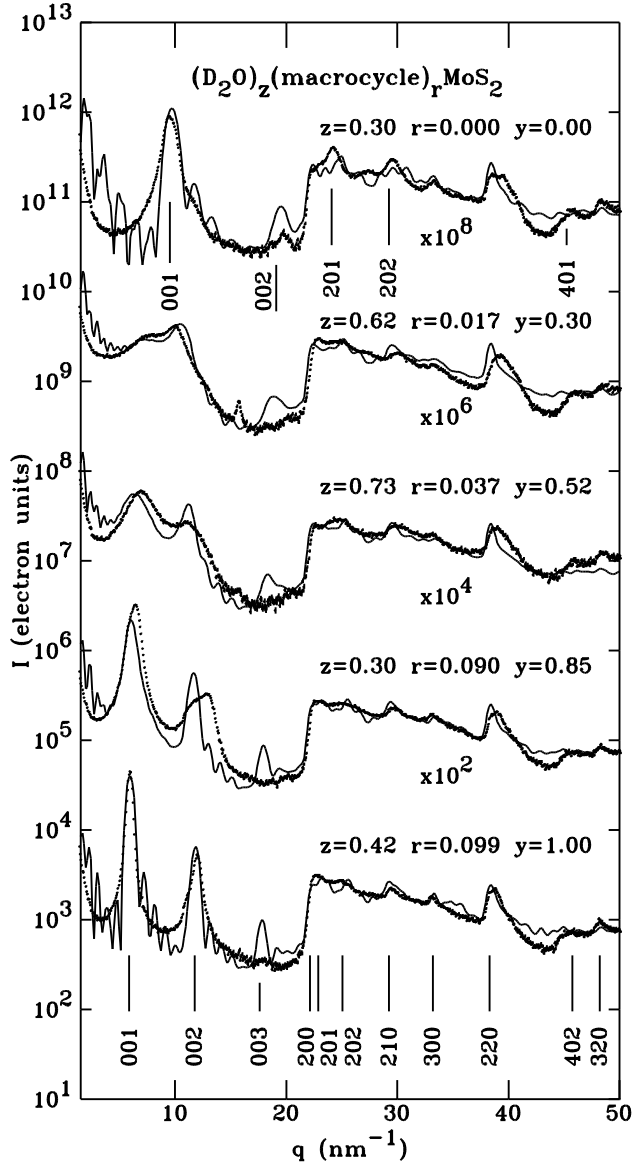}
\caption{\label{x-ray} Measured (points)  and simulated (lines)
X-ray scattering of intercalation compounds.
The fraction of macrocycles increases from top to bottom.  For clarity the data are multiplied by the indicated factors.  For the fully intercalated sample the discernible
peaks are indexed based on a hexagonal lattice with $a = 0.656 \rm \, nm$
and $c = 1.071 \rm \, nm$. For $r = 0$, the $l \ne 0$ peak positions correspond to
$c = 0.658 \rm \, nm$ with the same $a$. The fraction of open gallery spaces, $y$, has been
adjusted to provide the best fit to each spectrum.}
\end{center}
\end{figure}

Figure \ref{x-ray} shows the X-ray scattering of the intercalation compounds. 
The $(00l)$ peaks correspond to the stacking of the MoS$_2$ layers.  The step in the
scattering patterns near 22 nm$^{-1}$ is mostly due to the superposition of the
$(hk0)$ peaks that originate from the coherent scattering within the
MoS$_2$ layers.  The MoS$_2$ layers remain the same as
$r$ changes, with the exception of
the sample without intercalated macrocycles (restacked MoS$_2$, $r=0$).
We note that pristine MoS$_2$ has the 2H structure with trigonal prismatic coordination
beteen molybdenum and sulfur atoms, while exfoliated layers have the 1T structure with
octahedral Mo-S coordination \cite{yan,pae}.
For this $r=0$ sample the (201) and (202) peaks are enhanced.  This is
characteristic of the (partial) reconversion of the MoS$_2$ intra-layer
structure from the 1T to the
2H structure \cite{yan}.  It appears that the intercalated
macrocycles stabilize the octahedral Mo-S coordination by preventing
direct contact between MoS$_2$ layers. 

The sample with the highest content of intercalated macrocycles ($r = 0.099$) has a layer
spacing  of $1.06 \rm \, nm$, i.e.\ an  interlayer expansion of
0.45 nm relative to pristine MoS$_2$.  This expansion matches the thickness of the macrocycle,
and it agrees with the results for related intercalation compounds \cite{bis3}.
However, the measured $q$ positions of the $(00l)$ peaks  are not exact integer multiples as expected
for a regular lattice:  Lattice constants $c$ based on the
individual positions of the (001), (002) and (003) peaks are 1.057, 1.046 and 1.040 nm, respectively.
This may correspond to a modulation in the layer spacing (as discussed in Section \ref{modeling}).
Similarly the positions of the (001) and (002) peaks of the $r=0$ sample correspond to slightly
different lattice
constants $c$ (0.657 and 0.633 nm, respectively).
They are approximately reproduced by an overall spacing of 0.658 nm (0.043 nm interlayer expansion relative to
pristine MoS$_2$) and with closer spacings of some layers. 
Variations in amount of intercalated water are known to cause similar effects \cite{wes}.

The partially intercalated samples exhibit
broadening of the (001) and (002) peaks with decreasing macrocycle content.
Eventually the broadened peaks merge at the position of the narrower (001) peak of
restacked MoS$_2$. Thus the $(00l)$
peaks for the partially intercalated compounds are substantially broader than the peaks
for the $r=0$ and $r = 0.099$ samples.    An analogous broadening of
the peaks with decreasing guest species content
characterizes the spectra of MoS$_2$ pillard by hydroxyl-Al oligocations \cite{pei}.
This coexistence differs markedly from e.g.\ the behavior of clays with intercalated
surfactants.  For these clays the (001) peak shifts continuously with concentration,
indicating continuous swelling of the gallery spaces \cite{ghi}.
In section \ref{modeling} we reproduce the present data
with a model in which the MoS$_2$ layers are separated by random
sequences of open or closed gallery spaces.

Lamellar particles appear in scanning electron micrographs of restacked MoS$_2$ ($r=0$)  [figure
\ref{sem}(a)].  The  thinnest lamellae are about 40 nm thick.  
A micrograph of particles with open gallery spaces 
[figure \ref{sem}(b)] shows particles of similar diameter
that are covered by smaller lamellae that are about 200 nm wide.
The presence of smaller lamellae indicates that the guest molecules
make the particles more friable.
This may reflect overall weaker bonding between the MoS$_2$ sheets when the 
guest molecules are present.

\begin{figure}[t]
\begin{center}
\includegraphics{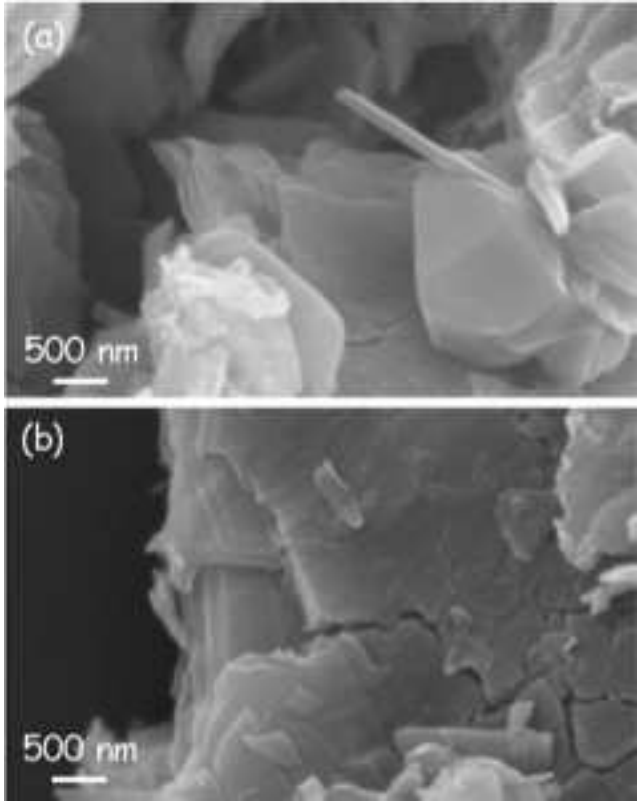}
\caption{\label{sem} Scanning electron microscope images of (a) restacked
(D$_2$O)$_{0.08}$MoS$_2$ particles and (b) intercalated particles
of composition (D$_2$O)$_{0.42}$(macrocycle)$_{0.099}$MoS$_2$.}
\end{center}
\end{figure}

Individual MoS$_2$ sheets can be seen in 
a TEM image of a cross-section of a particle of
(D$_2$O)$_{0.30}$(macrocycle)$_{0.090}$MoS$_2$.
The sheets are locally parallel, while overall
the particle has a mosaic structure with a range of different sheet orientations.
This is in agreement with TEM images of TiO$_2$-pillared MoS$_2$ \cite{pae}.
\begin{figure}[t]
\begin{center}
\includegraphics{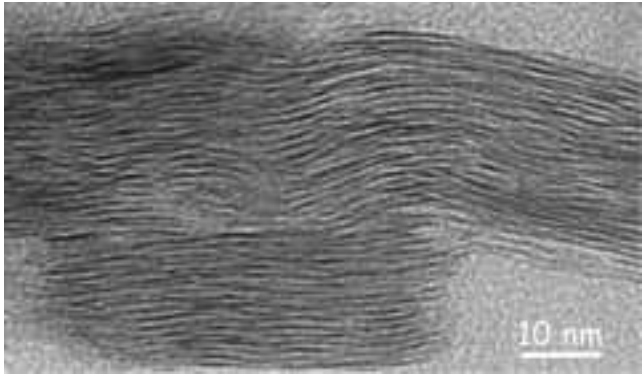}
\caption{\label{rb29detail}
TEM image of a particle of (D$_2$O)$_{0.30}$(macrocycle)$_{0.090}$MoS$_2$
showing the layer structure of the compound.
}
\end{center}
\end{figure}

The effect of increasing macrocycle concentration $r$ on the layer structure of the compound
is shown in Fig.\ \ref{tem_layers}.  The increase of the layer spacing with $r$ is obvious (all
images are to the same scale).  The restacked and the fully intercalated 
($r= 0.099$) compounds exhibit well ordered, parallel layers.  In contrast, the partially
intercalated compounds show a considerably higher degree of disorder.  Coexistence of
open and closed gallery spaces is seen most clearly in the $r = 0.037$ sample by the
presence of narrow and wide white spaces.  The
gallery spaces have regions that are locally open and closed, and transitions
between these regions are a significant part of the volume of the $r=0.017$
and $r=0.037$ samples.  As discussed below
in Section \ref{modeling}, this is accompanied by an approximate doubling
of the water content of these samples relative to the others.

\begin{figure}[t]
\begin{center}
\includegraphics{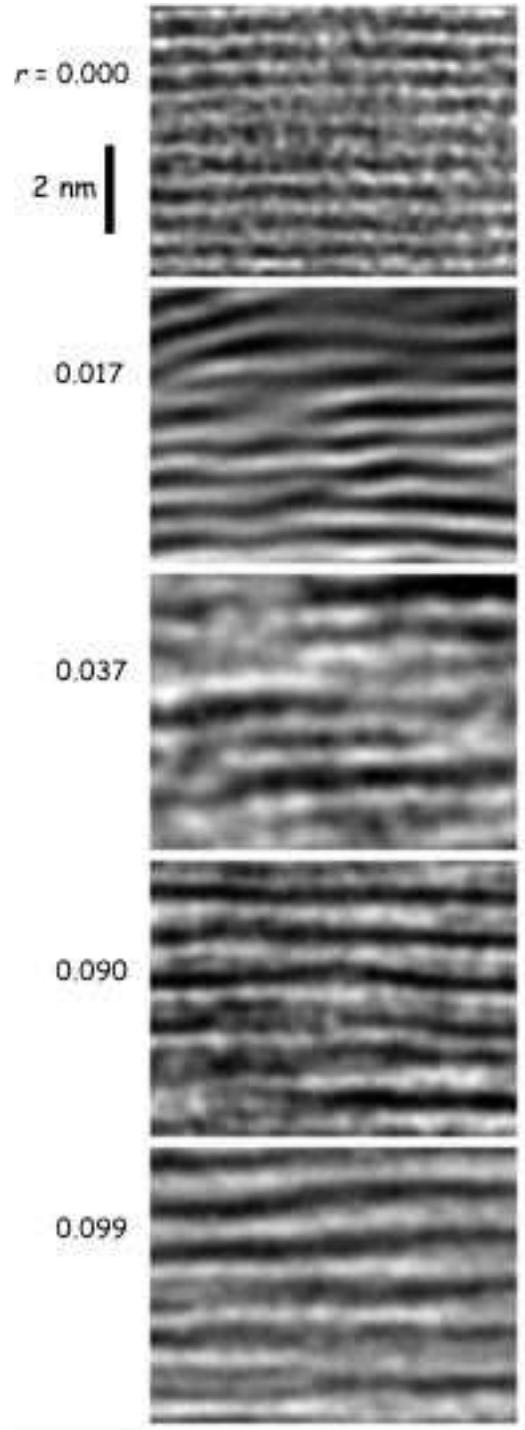}
\caption{\label{tem_layers}
Details of TEM images showing the stacking of the MoS$_2$ layers for
different macrocycle concentrations, $r$.
}
\end{center}
\end{figure}

Fourier transform power spectra of TEM images
of the layer structure are shown in figure \ref{fft} together with the X-ray scattering patterns
(logarithm of the scattered intensity).
Peak positions and intensity distributions of the $(00l)$peaks found with these two
different techniques agree well.
The Fourier transforms of
the images are not expected to show the step in intensity near $22 \, {\rm nm^{-1}}$ that
originates from the $h$, $k \neq 0$ peaks in the X-ray powder patterns. We conclude
that the stacking seen in the TEM images is indeed representative
of the bulk structure of  the samples.  

\begin{figure}[t]
\begin{center}
\includegraphics{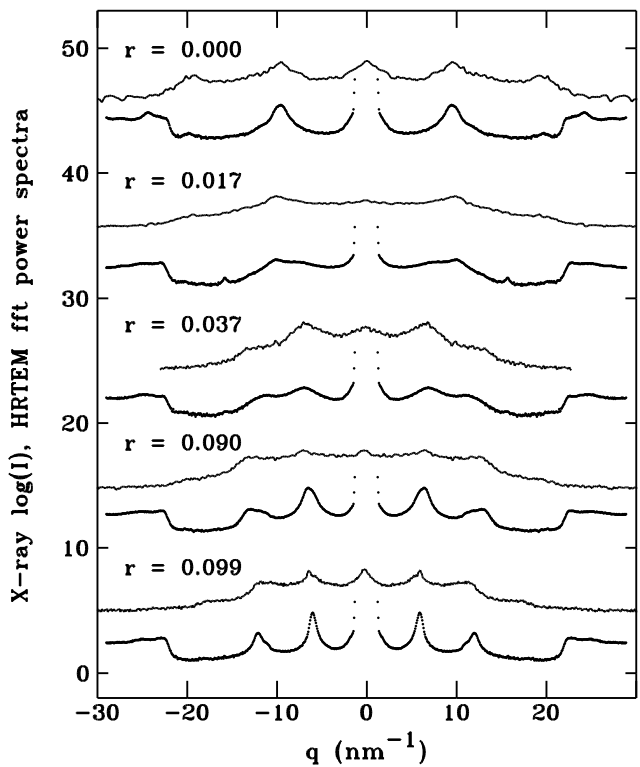}
\caption{\label{fft} Fourier transform power spectra of TEM images
(solid lines) and the logarithm of the X-ray scattering intensity (points)
for different concentrations of intercalated macrocycles, $r$.
For ease of comparison the X-ray data are plotted at positive and
negative $q$-values.  All intensities are in arbitrary units, and
curves are shifted by 10 units with decreasing $r$.}
\end{center}
\end{figure}

TEM images of the hexagonal basal planes of the compounds give similar
results for all $r$.
As an example, figure  \ref{tem_hex} shows the result for the fully intercalated compound. 
The hexagonal structure of the MoS$_2$
layers is evident.  Some disorder can be seen in the hexagonal grid structure,
e.g.\ groupings of of lattice points in pairs and triangles.  Random disorder
within the basal plane is a necessary feature in order to obtain good
agreement in modeling the X-ray scattering data (Section \ref{modeling}).

\begin{figure}[t]
\begin{center}
\includegraphics{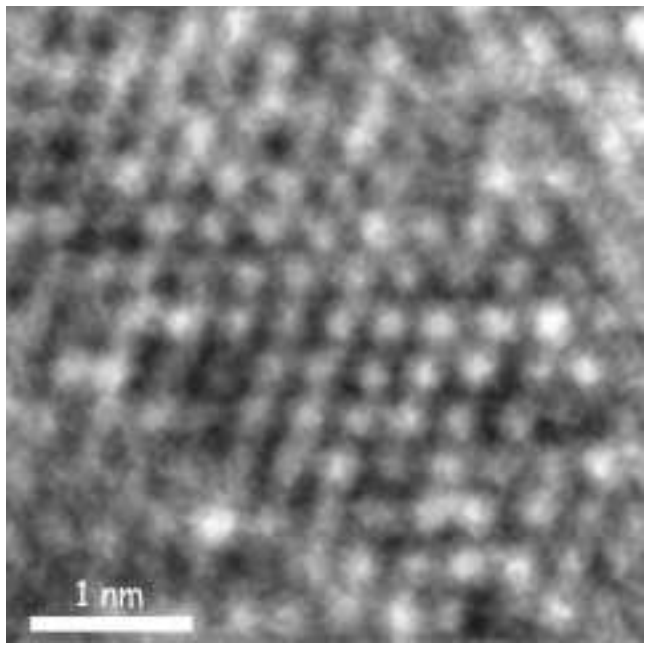}
\caption{\label{tem_hex}
TEM image of the basal plane of (D$_2$O)$_{0.42}$(macrocycle)$_{0.099}$MoS$_2$.
}
\end{center}
\end{figure}

The macrocycle binding enthalpy, $h_R$, is estimated from
the TGA data in the temperature range where desorbtion and evaporation
of the macrocycles from within the gallery spaces dominate the mass change.
The Clausius-Clapeyron equation relates the saturated vapor
pressure, $P$, and the temperature, $T$.  Considering the macrocycle vapor
and the absorbed macrocycles as the two coexisting phases, this relation is
\begin{equation}
\frac{d \ln P}{d (1/T)} = \frac{h_R}{k_B} ,
\end{equation}
where $k_B$ is Boltzmann's constant.
While the system is not in equilibrium during the thermogravimetric measurement,
we assume tentatively that the evaporative mass loss, ${-dm}/{d T}$, is proportional to
the number of molecules striking a container wall in equilibrium
({\it i.e.}\ that the powder sample acts as an effective Knudsen cell).
Time and temperature are proportional when the sample is heated at a constant rate.
The number of molecules incident on a container wall is
related to pressure and temperature via $PT^{-1/2}$ \cite{lan1,pri} so that
\begin{equation}
- \frac{d m}{d T} \propto P T^{-1/2}  .
\end{equation}
Hence plots of
$\ln \left[ T^{1/2} \left({-d m}/{d T}\right) \right]$ vs.\ ${1/T}$
should be straight lines with slope ${h_R}/{k_B}$.  These curves are plotted in
figure \ref{clausius} for the two samples with the highest macrocycle contents.
The data follow linear behavior
in the range corresponding to $100^{\circ} {\rm C} < T < 350^{\circ} {\rm C}$.
The slope corresponds to $h_R = -(0.14 \pm 0.03)$ eV per macrocycle.

\begin{figure}[t]
\begin{center}
\includegraphics{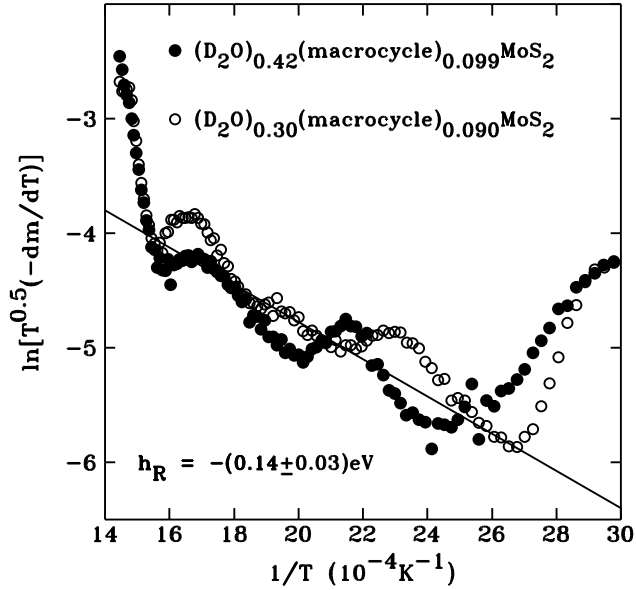}
\caption{\label{clausius} Scaled thermogravimetric data for the two samples with
the highest macrocycle contents.  The slope of the solid line gives
the approximate binding enthalpy per macrocycle absorbed in the gallery spaces, $h_R/k_B$.
}
\end{center}
\end{figure}

\section{Modeling the X-ray scattering}
\label{modeling}
Yang {\it et al.}\ found the structure of exfoliated single MoS$_2$ layers suspended in water by
comparing the X-ray scattering calculated for different models with the experimental
data \cite{yan,gor}.  We have extended the approach of comparing
calculated and measured diffraction patterns to MoS$_2$ intercalation compounds \cite{bru1}.    In an alternate approach, the structures of the ordered MoS$_2$ intercalation compounds containing metal atoms (Ag$_x$MoS$_2$ and LiMoS$_2$) have been solved using atomic pair distribution function analysis \cite{hwa,pet}.  This alternate approach matches the Fourier transform of diffraction data with the radial distribution of the atoms in the model.  This type of calculation has recently been applied to
diffraction data obtained with oriented films in enhance the precision of the results along
the direction perpendicular to the films \cite{wes1}.

Model particles representing correlated regions in the mosaic structure of the intercalation compounds
were built by placing the atoms according to a set of instructions listed below.
For each particle the X-ray scattering intensity was calculated according to the Debye formula
\begin{equation}
\label{debye_sum}
I(q)=  \sum_m {\sum_n { f_m f_n}} \sin {\left( q r_{mn} \right) } / {q r_{mn}} ,
\end{equation}
where $r_{mn}$ is the distance between atoms $m$ and $n$,
and $f_m$ is the atomic form factor of the $m$th atom \cite{yan,gor}.  Eq.\ (\ref{debye_sum})
allows for all possible orientations of the particles (powder average).  The simulated scattered intensity further includes factors that allow for the partial polarization of the
X-rays, the effect of the finite sample length at small angles (reflection geometry), and the incoherent scattering.  The simulated spectra were compared with the experimental result, and the instructions were modified in order to improve the agreement between calculation and spectrum by trial and error.

Here we consider a range of compounds.  They include restacked MoS$_2$
without intercalated macrocycles, fully intercalated MoS$_2$ with macrocycles in all gallery spaces,
and the partially intercalated samples.  The latter have some fraction $y$ of open gallery spaces containg
macrocycles, with the remainder being closed as in restacked MoS$_2$.
The experimental X-ray scattering patterns are reproduced by a model that assumes
that each sample is composed of powder particles that have random sequences of open
and closed gallery spaces.
Representing a closed
gallery space by a 0, and an open one with macrocycles by a 1,
a possible particle with seven layers can be represented by the six digit binary sequences
such as 001011.  We use equation (\ref{debye_sum}) to calculate the scattering of particles with all
possible combinations of open and closed gallery spaces.  This requires calculations for
36 of 64 binary sequences between
000000 and 111111, as the scattering of 28 sequences does not have to
be calculated due to the point inversion symmetry of equation \ref{debye_sum}.
One  of the model particles is displayed in figure \ref{stack}.
\begin{figure}[t]
\begin{center}
\includegraphics{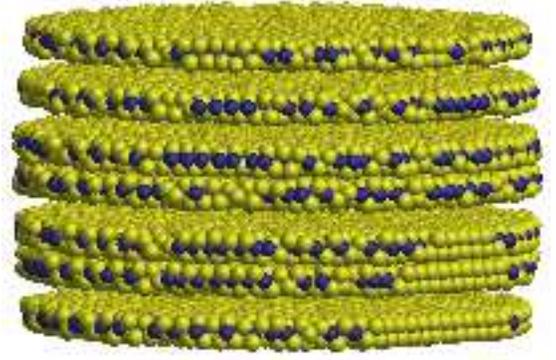}
\caption{\label{stack} Stacks of MoS$_2$ layers with open and closed gallery spaces between
the layers were generated in a computer simulation.  Shown is one of the 64 possible stacks consisting of seven MoS$_2$ layers.  The
example has four open gallery spaces.  The intercalated macrocycles occupying
open gallery spaces are not included in the model as they contribute little to the X-ray scattering.
}
\end{center}
\end{figure}
The scattering for samples with a given fraction of open gallery spaces, $y$, is obtained by 
averaging the scattering all model particles with the appropriate binominal weighting based
on the probability of
their sequence of open and closed gallery spaces.

Next the geometric properties of the model particles are described.
The information is organized according to length scales:
The atomic structure of a single MoS$_2$ layer in the $x$-$y$ plane,
the stacking of the layers and the finally
the shape and dimensions of the model particles.
 
\begin{enumerate}
\item
Structure of a single MoS$_2$ layer.
\begin{enumerate}
\item Underlying crystallographic structure of a single layer: Mo atoms are arranged in a hexagonal lattice, with the sulfur atoms arranged with octahedral symmetry, as described in
references [\onlinecite{yan,gor}].
\item \label{super} Superlattice: following reference [\onlinecite{yan}], a   superlattice with the Mo atoms at the positions (0.00, 0.00), (0.45, 0.00), (0.45, 0.00) and (0.45, 0.45) is introduced.
Other superlattice types have been considered \cite{gor,hwa,pet,hei}.
\item Lattice parameters: the lattice parameter of the $2 \times 2$ MoS$_2$  unit cell is 
$a = 0.656 \rm \, nm$.  The layer spacing in the stacking direction is  $c = 1.071 \rm \, nm$ for
an open gallery space and $0.658 \rm \, nm$ for a closed gallery space.
\item \label{dimer} Random dimerization: the distance between every pair of nearest neighbor Mo atoms is increased or decreased a random amount, with a maximum change of 12\%.
\item \label{trimer} Random trimerization: one third of the triangular groups of nearest
neighbor Mo atoms in the hexagonal lattice, selected at random, is contracted by 14\%, and one third is expanded by 14\%.
\item \label{sulfur} S atom positions: the sulfur atoms are placed above or below the center of mass of the three nearest Mo atoms at the distance 
$\left[ \left( 0.242 \, {\rm nm} \right)^2 -  {{\left( {d_1}^2+{d_2}^2+{d_3}^2 \right)} / 3} \right]^{1 / 2}$,
where $d_1$, $d_2$ and $d_3$ are the distances from the Mo atoms to the center of mass.
\end{enumerate}

\item
Stacking
\begin{enumerate}
\item Layer orientation: all MoS$_2$ layers have the same orientation within the $x$-$y$ plane.
\item \label{shifts} Layer shift: the layers are stacked with random shifts in the $x$-$y$ plane.
\item Intercalation: the intercalated molecules were omitted from the simulation of the
X-ray spectra, as they contribute little to the observed scattering \cite{bru1}.
\item \label{modulation} Layer spacing modulation: the spacing of the two outermost MoS$_2$ layers
of the model particles is reduced by 0.07 nm.
\end{enumerate}

\item
The particles
\label{particles}
\begin{enumerate}
\item Number of layers: There are seven MoS$_2$ layers and between zero and six open gallery
spaces.  All possible combinations of open and closed gallery spaces were generated.  The
calculated spectra are weighted averages of these spectra obtained by generating random
sequences with $(1-y)$ closed and $y$ open layers.  The model
particle with this sequence is then included in the average.
\item Size and shape: the outside shape of the model particle is a truncated rotational ellipsoid, with a maximum diameter of $10.2 \rm \, nm$  in the $x-y$ plane.
\end{enumerate}
\end{enumerate}

\begin{figure}[t]
\begin{center}
\includegraphics{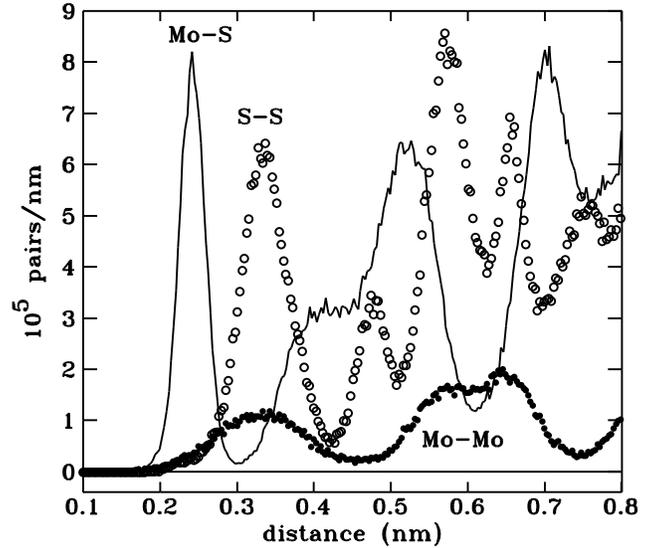}
\caption{\label{histogram} Partial pair distribution function for the model particle with all gallery spaces open ($y=1$).}
\end{center}
\end{figure}

These rules for the placement of the atoms were arrived at by a trial-and-error procedure that
was based on the comparison of the measured and modeled scattering intensities.
The influence of the different features assumed in the model on the simulated scattered
intensity has been discussed elsewhere \cite{bru1}.  The comparison of the placement rules
with the TEM pictures confirms some of the model features that were arrived at by the
trial-and-error fitting of the intensities.
The disorder present within the MoS$_2$ layer  (\ref{super}, \ref{dimer} and \ref {trimer} above)
is consistent with the features that can be seen in figure \ref{tem_hex}.  The random shifts in
stacking the layers (\ref{shifts}) correspond to the warping of the layers that can be seen in
figure \ref{rb29detail}.  The layer spacing moulation (\ref{modulation}) is required to redproduce
the slight irregularities in the position of the $(00l)$ peaks discussed in Section \ref{results}.

Modeled particles contain around 17500 atoms, so that evaluation of equation \ref{debye_sum} is computationally demanding.  Intensities at different $q$ (334 values) can be calculated in parallel. With 64 Xeon 2.4 GHz processors the calculation of the 36 required spectra takes about 200 min.  Figure \ref{histogram} displays the number of atom pairs
per unit length for different combinations of atom types for the model particle in which all gallery spaces are open.   The distribution of the Mo-Mo nearest neighbors
is broadened due to the disordering effects of the random dimerization and trimerization
of the Mo atoms, as well as the superlattice (\ref{super}, \ref{dimer} and \ref {trimer} above).
The Mo-S nearest neighbor distribution, based on procedure \ref{sulfur}, is considerably narrower.
The first peak in the S-S distribution includes nine near neighbors.
This peak is narrower than the distribution of
the six Mo-Mo nearest neighbors because in the model the sulfur atom positions are based on averages
of nearby Mo atom positions.
Since the sulfur atoms are closer to the (randomly) intercalated macrocycle molecules, the S-S coordination
in the real compound may be broader than the Mo-Mo distribution.

The simulated X-ray scattering patterns agree with the experimental data (figure \ref{x-ray}).  The
fraction of open gallery spaces, $y$, was adjusted to obtain the best match with the experimental data
for each sample.  The $y$'s for the best fits are given in figure \ref{x-ray}.
(Fitting the spectra of partially intercalated compounds to a linear combination of
particles with all open gallery spaces completely open or completely closed fails.)
The prominent broadening of the $(00l)$ peaks for partially intercalated
samples is, to a large extend, reproduced by the simulation.
However, the width of the simulated $(002)$ peak is insufficient
for the simulated $r = 0.037$ and $r = 0.090$  spectra.   Thus the model does not capture all the disorder present in partially intercalated compounds.
Specifically, the model disregards that within the same gallery space there are locally open and closed
regions.  This feature can be seen in figure \ref{tem_layers} e.g.\ for $r=0.017$ and $r=0.037$.
The calculation overestimates the (003) and (002) peaks near 19 nm$^{-1}$. 
Above $q = 20 \rm \, nm^{-1}$ the models match the measured scattering well.
The model does not take the partial transformation back to the trigonal prismatic 2H-MoS$_2$
structure into account that takes place only for restacked MoS$_2$ ($r=0$). 
In summary, the modeling procedure reproduces the experimental spectra
and it provides estimates for the fractions of open and closed gallery spaces.        

Fig.\ \ref{composition} shows that the fraction of open gallery spaces (determined by
matching the X-ray scattering data) increases approximately linearly with the macrocycle loading (determined by TGA).  For partial intercalation ($r=0.017$ and 0.037) the packing upon intercalation
appears to be less efficient, as the fraction of open gallery spaces is enhanced
and about twice the amount of water is cointercalated.  The series TEM images in
figure \ref{tem_layers} confirms this result.
\begin{figure}[t]
\begin{center}
\includegraphics{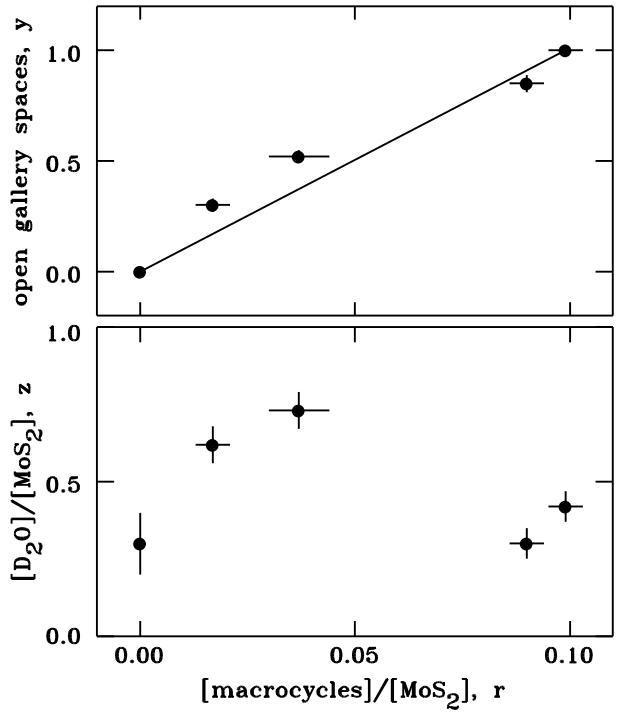}
\caption{\label{composition} Top: fraction of open gallery spaces as a function of
macrocycle loading.  The line is a guide to the eye.  Bottom: amount of intercalated water
as a function of macrocycle loading.}
\end{center}
\end{figure}

\section{Conclusions}
Macrocycle intercalation compounds with varying amounts of
intercalated molecules were prepared.  The sample compositions and the macrocycle binding enthalpy
were determined by thermogravimetric analysis.  TEM images show the evolution of
the host layer structure as the number of guest molecules increases.
The guest molecules intercalate as a single layer with the plane of  the macrocycle
parallel to the MoS$_2$ sheets.  Local regions of open and closed gallery spaces between
the MoS$_2$ layers coexist in samples that are not saturated with guest molecules. 
Within each layer there are transitions between open and closed regions that lead
to additional broadening of the $(00l)$ stacking peaks in the X-ray scattering.  A larger amount
of intercalated water is found in these partially intercalated samples.  We can be sure that
the TEM imaging process did not alter the stacking structure significantly
because Fourier transforms of the images agree with
the $(00l)$ peak intensities and positions found by X-ray powder diffraction of the bulk samples.

The X-ray scattering patterns of all compounds were reproduced by a set of
spatial models for the atom positions.
The models represent correlated regions within the compound that each contain about 17500 atoms.
These regions are
sufficiently small that their scattering can be evaluated directly as a Debye sum.
The model structures contain seven MoS$_2$ layers with all
possible sequences of open and closed gallery spaces,
and the scattering of all compounds was matched by a weighted averages of these scattering
patterns.  Only the fraction of open gallery spaces in the weighted average
is adjusted to match composition dependence of the X-ray scattering patterns.
The model MoS$_2$ layers stack with the same orientation,
but with random lateral shifts.  The models as well as TEM images
of the basal plane show that the MoS$_2$ layers are highly disordered.  Compared to this disorder
periodic superlattice structures are of lesser importance.

\section{Acknowledgements}
The scanning electron microscopy sample preparation and imaging was carried out by Jim Ehrman
(Mount Allison Digital Microscopy Facility).  Transmission electron
microscopy sample preparation and imaging were
carried out by Louise Weaver (UNB  Microscopy and Microanalysis).
The X-ray modeling calculations were performed
with the Mount Allison Unviersity computer cluster (Torch).
The Natural Science and Engineering Research Council of Canada has supported this project
with Discovery Grants (Bissessur, Haines and Br\"uning).
Bissessur gratefully acknowledges Canada Foundation for
Innovation, Atlantic Innovation Fund, and University of Prince
Edward Island for financial support.


\bibliography{intercalation}

\end{document}